\documentclass[conference]{IEEEtran}
\usepackage{amsmath}
\usepackage{amsfonts}
\usepackage{amssymb}
\usepackage{cite}
\usepackage{multicol}
\usepackage{verbatim}
\usepackage{graphicx}
\usepackage{balance}
\usepackage{algorithm}
\usepackage{algpseudocode}
\usepackage{epstopdf}
\usepackage{epsfig,color,algorithm}

\IEEEoverridecommandlockouts
\graphicspath{ {D:/j-images/} }

\begin{document}

\title{Enhancing Physical Layer Security in AF Relay Assisted Multi-Carrier Wireless Transmission}
\author{Waqas Aman$^1$, Guftaar Ahmad Sardar Sidhu$^2$, Haji M. Furqan$^3$, Zain Ali$^4$\\
$^1$Electrical Engineering Department, Information Technology University, Lahore, Pakistan\\
$^{2,4}$Department of Electrical Engineering, COMSATS Institute of Information Technology, Islamabad, Pakistan\\
$^3$School of Engineering and Natural Sciences, Istanbul Medipol University, Istanbul, Turkey\\
Emails: $^1$waqas.aman@itu.edu.pk, $^2$guftaarahmad@comsats.edu.pk,
\\$^3$hamadni@st.medipol.edu.tr, $^4$zainalihanan1@gmail.com \\
}

\maketitle 


\begin{abstract}
In this paper, we  study the physical layer security (PLS) problem in the dual hop orthogonal frequency division multiplexing (OFDM) based wireless communication system. First we consider a single user single relay system and study a joint power optimization problem at the source and relay subject to individual power constraint at the two nodes. The aim is to maximize the end to end secrecy rate with optimal power allocation over different sub-carriers. Later, we consider a more general multi-user multi-relay scenario. Under high SNR approximation for end to end secrecy rate, an optimization problem is formulated to jointly optimize power allocation at the BS, the relay selection, sub-carrier assignment to users and the power loading at each of the relaying node. The target is to maximize the overall security of the system subject to independent power budget limits at each transmitting node and the OFDMA based exclusive sub-carrier allocation constraints. A joint optimization solution is obtained through duality theory. Dual decomposition allows to exploit convex optimization techniques to find the power loading at the source and  relay nodes. Further, an optimization for power loading at relaying nodes along with relay selection and sub carrier assignment for the fixed power allocation at the BS is also studied. Lastly, a sub-optimal scheme that  explores joint power allocation at all transmitting nodes for the fixed sub-carrier allocation and relay assignment is investigated. Finally, simulation results are presented to validate the performance of the proposed schemes.

\end{abstract}

\section{Introduction}
 
\IEEEPARstart{D}ual- hop communication has recently gained significant attention in the field of wireless communication due to its better performance over single-hop communication\cite{22}. In dual-hop communication a relay is used as an intermediate node between sender and receiver. It is generally used to enhance throughput, reduce power consumption, and to increase coverage area at the cell edges. There are two types of relaying protocols that are widely used: Amplify-and-Forward (AF) and Decode-and-Forward (DF). The AF relaying protocol first receives signal from the source and then forwards it to the destination with amplification, while DF relaying protocol first receives signal from the source, decodes it, re-encodes it and then forwards the resultant signal to the destination\cite{44}.

The broadcast nature of wireless communication provides many exciting opportunities, however it makes the security of link a challenging issue. Wireless communications can potentially be attacked by malicious nodes, and therefore, security issues have taken an important role in today's communications \cite{FGAO}. A promising technique for achieving secure communications is Physical Layer Security (PLS) \cite{n5}. A wireless link is considered to be secure if it provides a positive non zero secrecy rate and a link with  higher secrecy rate is known to be more secure link \cite{a}. 

To provide PLS in dual-hop single carrier networks, resource allocation has been widely studied under DF relaying protocol  \cite{2}--\cite{8n}. The authors in\cite{2} and\cite{3} studied the problem of optimal relay placement to enhance PLS.
The work\cite{6} considered joint relay selection and power optimization to maximize the system's secrecy rate. Further, \cite{7n} proposed a joint relay and jammers selection with power optimization. Recently, \cite{8n} discussed the the relay selection in the presence of adaptive eavesdropper. The dual-hop transmission under amplify and forward (AF) protocols has become much attractive due to its simple implementation\cite{z7}. However, the resource allocation in AF relay enhanced networks has always been a challenging task\cite{z88}. Different aspects of PLS in AF based single carrier systems has been studied in \cite{Huang}--\cite{9n}. The authors in\cite{Huang} investigated the impact of using an untrusted AF relay on secure communication and derived the exact Secrecy Outage Probability (SOP) under different transmission scenarios.  With mulitple trusted relays,\cite{LFan} proposed different relay selection strategies to enhance the PLS in multi-user cooperative relay networks. The work in \cite{Akhtar} focused on achievability of secrecy rate under different channel conditions. 
The sub optimal relay selection with fairness is studied in \cite{8nn}. The relay transmit power optimization protocols for secrecy maximization under both AF and DF has been studied in\cite{9n}.

The multi-carrier transmission has become a fundamental choice for the next generation wireless communication networks because of its ability to combat multi path fading effects, high spectral efficiency, and provision of flexibility in resource allocation\cite{1}. To provide PLS in multi-carrier systems, resource optimization is one of the popular technique and has been studied in\cite{14}-\cite{18}. In\cite{14} a dynamic sub-carrier allocation for secure transmission is studied in the presence of passive eavesdropper. The proposed scheme utilizes the Channel State Information (CSI) between legitimate users and drops out highly faded sub-carrier and modifies the modulation scheme for remaining good sub-carriers, to achieve better secrecy rate. Further, \cite{15n} provided a optimal sub-carrier allocation for outage probability minimization with secrecy constraint. In \cite{16n}, authors proposed the optimal power allocation with sub-carrier allocation to maximize the sum rate with outage probability and fairness constraints. In\cite{17}, two categories of users were considered: the secure users and the non secure users. The task was to maximize the throughput of non-secure users via optimal power allocation subject to guaranteed average secrecy rate to secure users. The work in\cite{active} extended the previous work to maximize secrecy rate in the presence of active eavesdropper which has the capability to jam the secret user transmission. In the presence of both active and passive eavesdroppers, the power allocation over sub-carriers to maximize the average secrecy rate has been investigated in\cite{17d}. Further, the authors in \cite{18} considered multiple eavesdroppers and optimized sub-carrier assignment, power allocation, and secrecy data rate to maximize the energy efficiency.

\subsection{Related Work and Contributions}
 Under the umbrella of OFDMA, resources allocation for PLS in dual hop with DF protocol has been studied in \cite{4}--\cite{df2016}. Power allocation problem to maximize the secrecy rate was investigated in\cite{4}. The extension to this work with joint sub-carrier allocation and power loading was made in\cite{5}. Recently, \cite{df2016} proposed two different power optimization schemes at source and relay under individual power constraint, one achieves sum secrecy rate maximization in the presence of untrusted users while the other achieves fairness for minimum requirement of secrecy per user. The dual hop communication under DF relaying is allowed for the trusted relaying nodes only. However, if the relay node is un-trusted, AF protocol becomes the better choice as it does not require any decoding at the relay. However, the resource allocation schemes designed to enhance PLS under DF transmission can not be directly applied to AF scenario. 
 
Recently, different works on PLS in dual hop systems under AF relaying protocol have been reported \cite{Jendal}--\cite{jendal2016}. The resource optimization in Orthogonal Frequency Division Multiplexing (OFDM) based single-user single-relay systems was considered in \cite{Jendal}. The authors studied the sub-carrier assignment and power allocation strategies under a total system power constraint. The optimization under sum power constraint provides a good analysis of power allocation, however it may not be an attractive solution for practical systems. Further, in \cite{lat} the authors investigated the power allocation at source node and sub-carriers allocation among users in a single  relay multiuser system. Recently, \cite{waqas} extended the work to  multi-relay scenario and considered the relay assignment and power allocation problem. However, both the works in \cite{lat} and \cite{waqas} considered power allocation at the source node only while the power optimization at the relaying node(s) was missing. The power optimization only at the source node simplifies the solution at the cost of degradation in performance. More recently, \cite{jendal2016} investigated the power allocation at the source and the relay nodes under a single user single relay scenario. The authors proposed a sub-optimal solution through  alternate optimization approach. A joint optimization of power allocation at the source and the relay nodes along with sub-carrier assignment and  relay selection can provides much more benefits. This joint optimization is a challenging task and to the best of authors' knowledge has not been investigated yet. 

In this work, our aim is to maximize the sum secrecy rate under AF relaying protocol in single cell down-link transmission. We first consider the joint power allocation at the source and the relay nodes subject to separate power constraint at each node. The end to end secrecy rate under AF protocol depends on both hops, i.e., the power allocation at the two nodes is coupled with each other. Thus, instead of separate power optimization \cite{jendal2016} at the source and relay, we propose a joint optimization solution. Then, we consider a joint sub-carrier allocation, relay selection, and power allocation problem in a multi-user multi-relay system. Various solution schemes are proposed to efficiently solve the problem.   Our contributions are summarized as:
\begin{itemize}
\item We solve a joint power allocation problem in an OFDM based dual hop network to optimize the power distribution among different sub-carriers at the source and the relay node. An efficient solution is obtained through Karush-Kuhn-Tucker (KKT) optimality conditions. 
\item Later, a novel joint optimization problem is formulated which considers the power allocation at the source node, the optimal relay assignment to users, the sub-carrier allocation to each assigned relaying node, the power allocation at each relay node, and the sub-carrier allocation to each user subject to separate power constraint at the source and each of the relaying node.      
\item A joint solution of the  mixed integer programming problem is obtained through efficient dual decomposition techniques to maximize the overall system's secrecy rate. 
\item To look into the effect of power optimization at the relaying nodes only, we redefine the joint problem for uniform power allocation the source node and similar techniques are used to solve this problem. 
\item Finally, a low complexity sub-optimal algorithm is proposed which optimizes the power at the source and the multiple relaying nodes for the predefined sub-carrier allocation and the relay assignment. 
\item Extensive simulation results are presented to evaluate the performance of the proposed schemes.    
\end{itemize} 

The remainder of this paper is organized as follows.   The joint power allocation at the source and the relay node in a single user single relay case is presented in Section-\ref{singleuser}. Under multi-user multi relay system, the proposed framework is elaborated in Section-\ref{jnt}. The Section-\ref{s3b} includes proposed solution for power allocation at the relaying nodes along with sub-carrier allocation and the relay assignment, where the problem of power allocation at the source and the multiple relaying nodes without optimizing other parameters is considered in Section-\ref{s3c}. Finally, the simulation results and the conclusion are presented in Section-\ref{sim} and Section-\ref{con}, respectively.

\section{Joint Power Allocation at the Relay and the Source Node}\label{singleuser}
\subsection{System Model and Problem Formulation}\label{sys1}
In this section, we consider a dual hop multi-carrier system which consists of a source node (S), an AF relay  node (AR), a destination node (D), and an eavesdropper (Eve) as shown in Fig. 1.
\begin{figure}[h]
\includegraphics[width=9cm, height=7cm]{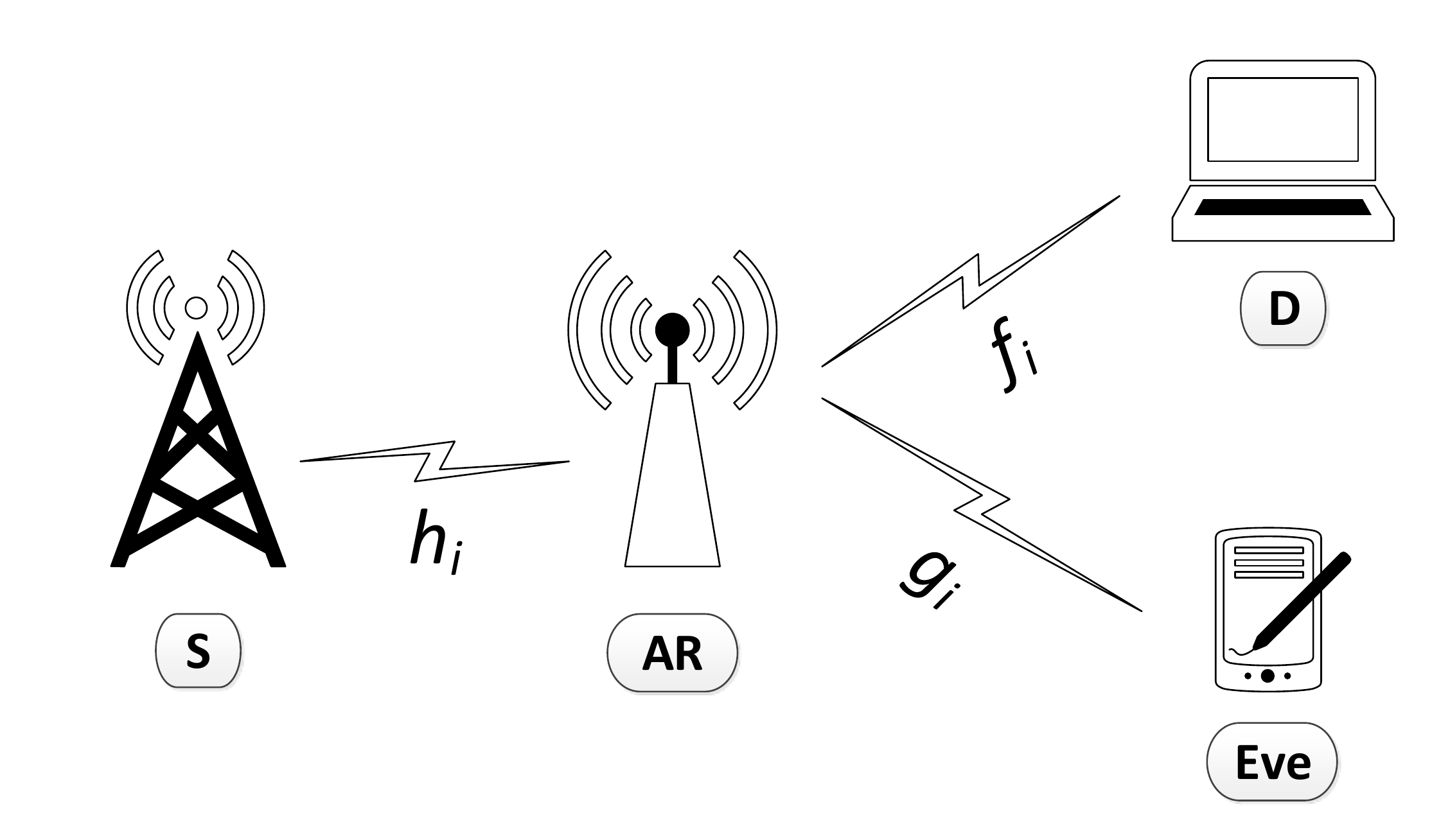}
\caption{OFDM based Single User Single Relay System}
\end{figure}
We assume that all devices are equipped with single antenna and D and Eve are co-located  such that the direct path from S to D and S to Eve is missing due to large distance [27], [29], [30]. The channel gains of $i$-th sub-carrier over S-to-AR, AR-to-D, and AR-to-Eve links are denoted by $h_i$, $g_i$, and $f_i$, respectively. In the first transmission slot, the AR receives a message signal over $i$-th sub-carrier and re-transmits with amplification factor $Q_i$, given by
\begin{align}
Q_i= \sqrt{\frac{q_i}{p_i|h_i|^2+\sigma^2}},
\end{align}  
where $p_i$ and $q_i$ are the power loading over $i$-th carrier at S and AR, respectively, and $\sigma^2$ denotes the  variance of Additive White Gaussian Noise (AWGN).
\begin{table*} [t] 
 \begin{center}
     \caption{}
 \begin{tabular}{|c|c|}
   
   \hline  $A_i$ & 
    $H_i^2\lambda(H_iG_i^3V-H_iG_iF_i^2V-H_iG_i^2F_iV+H_iF_i^3V
   $\nonumber\\&\qquad  $+G_i^3F_i\lambda+G_iF_i^3\lambda-2G_i^2F_i^2\lambda)$ \\ 
   \hline $B_i$	& $(2H_i^3G_i^2+2H_i^3F_i^2-4H_i^3G_iF_i-2H_i^2G_i^2F_i\lambda V+3H_i^2G_i^3\lambda V-3H_i^2G_iF_i^2\lambda V+H_i^2F_i^3\lambda V$
    \nonumber\\&\qquad $+2H_iG_i^3F_i\lambda^2+2H_iG_iF_i^2\lambda^2-4H_iG_i^2F_i^2\lambda^2-H_i^3G_i^3V-3H_i^3G_iF_i^2V+3H_i^3G-i^2F_iV+H_iF_i^3V)$ \\
     \hline $C_i$& $(H_i^2G_i^2+H_i^2F_i^2-2H_i^2G_iF_i+H_iG_i^3\lambda V-2H_iG_i^2F_i\lambda V+H_IG_IF_i\lambda V-2H_iG_iF_i^2\lambda V+G_i^3F_i\lambda^2$
     \nonumber\\&\qquad $+G_iF_i^3\lambda^2-2G_i^2F_i^2\lambda^2-H_i^2G_i^3V-3H_i^2G_iF_i^2V+3H_i^2G_i^2F_iV+H_i^2F_i^3V)$ \\ 
   \hline
   \end{tabular} 
   \end{center}
   \end{table*}

The received signal-to-noise ratio (SNR) at the D over $i$-th sub-carrier can be expressed as
\begin{align}
SNR_i^{D}= \frac{Q_i^2p_i|h_i|^2|g_i|^2}{Q_i^2|g_i|^2\sigma^2+\sigma^2 }.
\end{align}  
Similarly, received SNR at the Eve is given as
\begin{align}
SNR_i^{E}= \frac{Q_i^2p_i|h_i|^2|f_i|^2}{Q_i^2|f_i|^2\sigma^2+\sigma^2}.
\end{align}  
The secrecy rate over $i$-th sub-carrier can be expressed as
\begin{align}
\text{SR}_i= \log_2\left(1+SNR_i^{D}\right)-\log_2\left(1+SNR_i^{E}\right).
\end{align}
Let $N$ be the total number of sub-carriers,  the sum secrecy rate under high SNR approximation can be written as \cite{z8}:
\begin{align}
\text{SR}_{sum}= \frac{1}{2}\sum_{i=1}^N \underbrace{\log_2\left(\frac{G_i+H_iG_ip_i+G_iF_iq_i}{F_i+H_iF_ip_i+G_iF_iq_i}\right)}_{SR_i},
\end{align}
where $H_i=\frac{|h_i|^2}{\sigma^2}$, $G_i=\frac{|g_i|^2}{\sigma^2}$, $F_i=\frac{|f_i|^2}{\sigma^2}$, and the term $\frac{1}{2}$ appears due to half duplex relay transmission.

Our target is to maximize the sum secrecy rate of the system by optimizing power over the sub-carriers at the S as well as at the AR under individual power constraints. Thus, the optimization problem becomes
\begin{align}
 \max_{p_{i},q_{i}}\quad & \sum_{i=1}^NSR_i  \label{opt1}\\
 & \sum_{i=1}^Np_i\leq P_t, \quad \sum_{i=1}^Nq_i\leq Q_t.  \nonumber
\end{align}
The first constraint ensures that the total power allocated to the all sub-carriers at S must be within the total available power $P_t$. Similarly, the second constraint ensures that the allocated power over all sub-carriers at AR node should not exceed the maximum limit $Q_t$.
\subsection{Proposed Optimization Scheme}\label{s2b}
The problem (\ref{opt1}) is a convex optimization problem and we use the duality theory to obtain the solution. The optimal power loading can be obtained from the following dual problem 

\begin{align}
\min_{\lambda\geq0,V\geq0}\ \ &\max_{p_i\geq0,q_i\geq0} \sum_{i=1}^NSR_i + \lambda\left(P_t- \sum_{i=1}^Np_i\right)\nonumber \\&\quad +V\left(Q_t-\sum_{i=1}^Nq_i\right),
\end{align}  
where $\lambda$ and $V$ are the associated dual variables. Removing the constant terms, the problem can be re-written as 
\begin{align}
\min_{\lambda\geq0,V\geq0} \ \ &  \sum_{i=1}^N \max_{p_i\geq0,q_i\geq0} \log_2\left(\frac{G_i+H_iG_ip_i+G_iF_iq_i}{F_i+H_iF_ip_i+G_iF_iq_i}\right)\nonumber \\& - \lambda p_i-V q_i. \label{7}
\end{align} 
Applying KKT conditions to the internal maximization, we obtain
\begin{align}
p_i^*= \left(\frac{-B_i+\sqrt{{B_i}^2-4A_iC_i}}{2A_i}\right)^+,\label{opt_p}
\end{align}
and
\begin{align}
q_i^* = \left(\frac{\lambda D_i}{VE_i}+\frac{\lambda B_i+\sqrt{(\lambda B_i)^2-4A_iC_i\lambda^2}}{2VA_i}\right)^+ \label{opt_q},
\end{align}
where $(x)^+=\max(0,x)$, $D_i=\left(F_i-G_i\right)$, $E_i=\left( G_iH_i-H_iF_i\right)$ and the values of $A_i$, $B_i$ and $C_i$ are given by Table 1.

The problem in (\ref{7}) becomes: 
\begin{align*}
\min_{\lambda\geq0,V\geq0} \ \ &  \sum_{i=1}^N \log_2\!\left(\frac{G_i+H_iG_ip_i^{*}+G_iF_iq_i^{*}}{F_i+H_iF_ip_i^{*}+G_iF_iq_i^{*}}\right)\! -\! \lambda p_i^{*}\!-\!V q_i^{*}.
\end{align*} 
 
To find the dual variables, we use the following iterative sub-gradient updates\cite{gd}--\cite{9}
 \begin{align}
\lambda(m)=\lambda(m-1)+\delta\left(P_t-\sum_{i=1}^Np_i^* \right), \label{10}
\end{align}
 \begin{align}
V(m)=V(m-1)+\delta\left(Q_t-\sum_{i=1}^Nq_i^*\right), \label{11}
\end{align}
where $m$ represents the $m$-th iteration and $\delta$ is the step size. In each update of dual variables, the optimum power allocation at the BS and the relay are obtained from (\ref{opt_p}) and (\ref{opt_q}). At convergence, the optimum values of dual variables as well as of power variables are obtained. 

\section{Joint sub-carriers allocation, relays selection and power allocation}\label{jnt}
In this section, we consider multi-user, multi-relay and multi-carrier dual hop communication with a single base station (BS), $K$ number of  secret users, $J$ number of AF relays (ARs), $N$ numbers of sub-carriers and a single Eve as shown in Fig. 2.
The channel gain from BS to j-th AR node on i-th sub-carrier is denoted by $h_{i,j}$. The channel gain from j-th AR node to k-th user on i-th sub-carrier is denoted by $g_{i,j,k}$, the corresponding channel gain from j-th AR node to Eve is denoted by $f_{i,j}$ and $u_{i,j}$ is the power allocated over the i-th sub-carrier at the j-th relay. With this, the secrecy rate over $i$-th sub-carrier at the the $k$-th user communicated through $j$-th relay can be expressed as: 
\begin{align}
\text{SR}_{i,j,k}=&\frac{1}{2}\log_2\left(\frac{{b_{i,j,k}+a_{i,j}b_{i,j,k}p_i+u_{i,j}b_{i,j,k}c_{i,j}}}{{c_{i,j}+a_{i,j}c_{i,j}p_i+u_{i,j}b_{i,j,k}c_{i,j}}}\right),
\end{align}
where $a_{i,j}=\frac{|h_{i,j}|^2}{\sigma^2}$, $b_{i,j}=\frac{|g_{i,j,k}|^2}{\sigma^2}$, and $c_{i,j}=\frac{|f_{i,j}|^2}{\sigma^2}$.
We adopt a fully flexible AR allocation strategy where a relaying node can be allocated to more than one users, and each user can be served with multiple AR nodes over different sub-carriers. Furthermore, a sub-carrier is allocated to the same user over the  two hops of transmission\footnote{The information received over $i$-th sub-carrier at the first hop can be forwarded over a different carrier in the second hop, however is beyond the scope of this work}. On account of sub-carrier allocation and AR selection, we define two binary  variables:  $\alpha_{i,k} \in {[0,1]}$ such that $\alpha_{i,k}=1$ when the $i$-th sub-carrier is allocated to the $k$-th user and zero otherwise, and $\beta_{j,k} \in {[0,1]}$ such that $\beta_{j,k}=1$ when the $j$-th AR is allocated to the $k$-th user and zero otherwise. With this, the sum secrecy rate of the system can be expressed as;
\begin{align}
\text{SR}_{\text{sum}}=\sum_{i=1}^N\sum_{j=1}^J\sum_{k=1}^K \alpha_{i,k}\beta_{j,k} \text{SR}_{i,j,k}.
\end{align}
\begin{figure*}
\includegraphics[width=18cm, height=7cm]{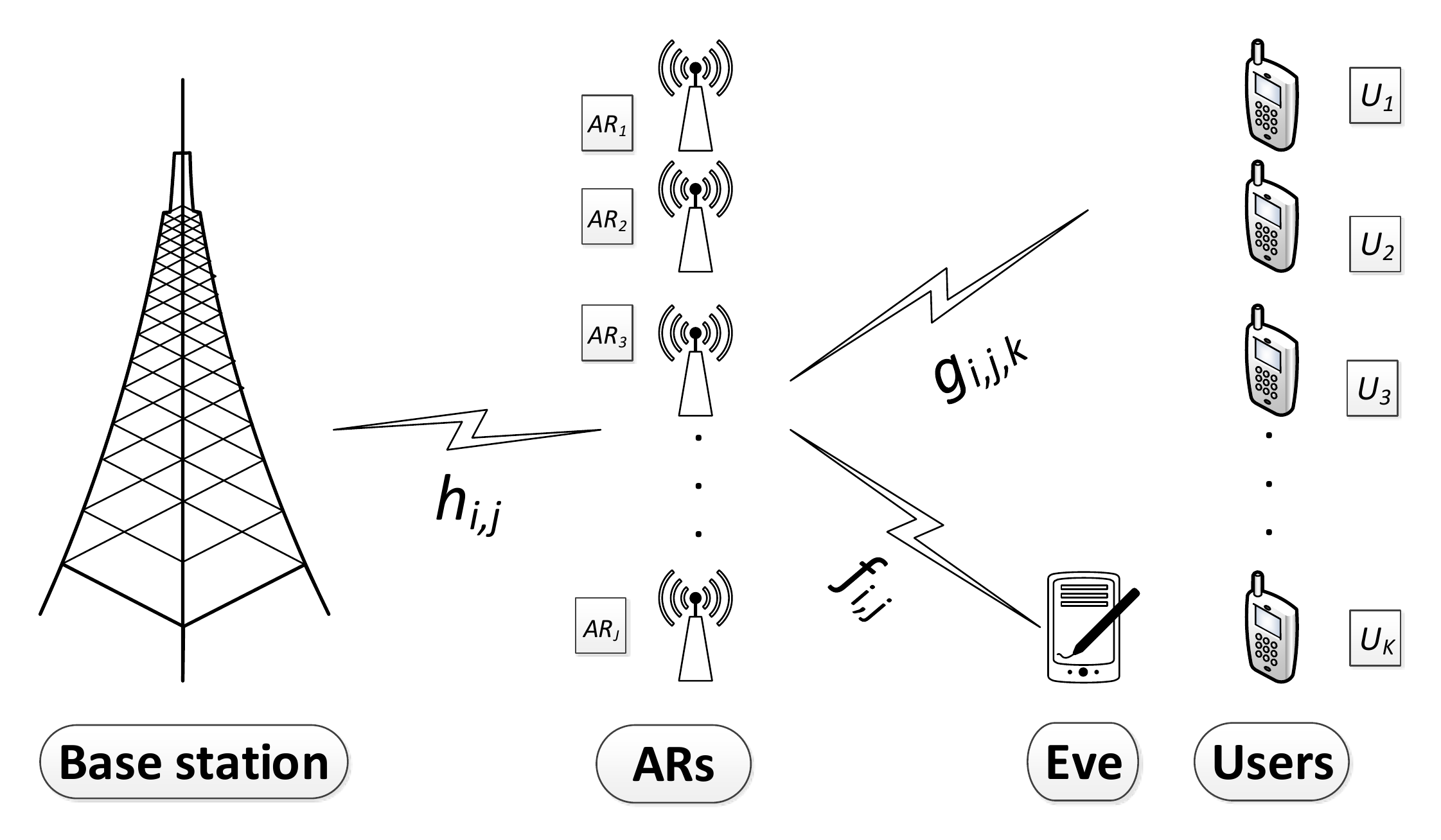}
\caption{OFDMA based Multi-User Multi-Relay System}
\label{Fig 3: Sum Secrecy Rate vs Total Transmit Power (PT)}
\end{figure*}

\subsection{Problem Formulation}
The aim is to maximize SR$_{\text{sum}}$ with jointly optimizing the AR assignment, sub-carrier allocation, BS's transmit power loading  and  power allocation at the relaying nodes over different sub-carriers. Let $P_t$ and $Q_{t,j}$ be the total powers available at BS and j-th AR, respectively. Then, the joint sub-carrier allocation, AR assignment, and power loading optimization can be formulated as:
\begin{align}
\max_{(p_{i},u_{i,j},\alpha_{i,k},\beta_{j,k})}\quad &\ \text{SR}_{\text{sum}}\label{18} \\
\text{s.t.} \ \ \quad &\sum_{k=1}^K\alpha_{i,k}= 1, \nonumber \quad \forall i,
\\&\sum_{i=1}^N\sum_{j=1}^J\sum_{k=1}^K \alpha_{i,k}\beta_{j,k}p_i \leq P_t, \nonumber
\\&\sum_{i=1}^N\sum_{k=1}^K \alpha_{i,k}\beta_{j,k}u_{i,j} \leq Q_{t,j} \;\;\forall j. \nonumber
\end{align}
The first constraint ensures that a particular sub-carrier can not be assigned to more than one users. The second constraint represents that the sum transmit power on all sub-carriers at BS should be less than or equal to a maximum power limit $P_{t}$ and the last constraint guarantees that total transmit power over different sub-carriers at the j-th AR should be less than or equal to a maximum power budget $Q_{t,j}$.
\subsection{Proposed Solution}
The problem (\ref{18}) is a  mixed binary integer programming problem, and a vast search over all variables is needed to find an optimal solution. Thanks to \cite{8}, the difference between the solution of dual problem and the solution of primal problem\footnote{Commonly known as duality gap.}  becomes zero when we have sufficiently large number of sub-carriers in OFDM based transmission regardless of convexity of original problem. The dual problem associated with primal problem (\ref{18}) can be defined as:
\begin{align}
\min_{\left(\lambda\geq0 , \; V_j\geq0\right)} \quad D(\lambda, V_j)\label{19},
\end{align}
where $\lambda$ and $V_j$ are the dual variables, and the dual function $D(\lambda, V_j)$ can be expressed as
\begin{align}
D(\lambda,V_j)=&\max_{(p_{i}, u_{i,j}, \alpha_{i,k},\beta_{j,k})}  L(p_{i}, u_{i,j},\alpha_{i,k},\beta_{j,k})\label{20}\\\nonumber \\
\text{s.t.}\ \ \quad &\quad\sum_{k=1}^K \alpha_{i,k} = 1 \nonumber ,\quad \forall i,
\end{align}
with,
\begin{align}
&L(p_{i}, u_{i,j},\alpha_{i,k},\beta_{j,k})=\\&\sum_{i=1}^N\!\sum_{j=1}^J\!\sum_{k=1}^K\! \alpha_{i,k}\beta_{j,k} \text{SR}_{i,j,k}\!+\!\lambda\!\left(\!P_t\!-\!\sum_{i=1}^N\sum_{j=1}^J\sum_{k=1}^K \alpha_{i,k}\beta_{j,k}p_i\!\right) \nonumber
\\&+\sum_{j=1}^JV_j\left(Q_{t,j}-\sum_{i=1}^N\sum_{k=1}^K \alpha_{i,k}\beta_{j,k}u_{i,j}\right). \nonumber
\end{align}
To solve the dual problem we first solve the dual function $D(\lambda,V_j)$ and  similar to \cite{9} we adopt dual decomposition approach. The problem in (\ref{20}) can be rewritten as:
\begin{align}
D(\lambda, V_j)=&\max_{(p_{i}, u_{i,j},\alpha_{i,k},\beta_{j,k})} \sum_{i=1}^N\sum_{j=1}^J\sum_{k=1}^K \alpha_{i,k}\beta_{j,k}(SR_{i,j,k},\label{22}\\&-\lambda p_i-V_ju_{i,j})\nonumber +
 \lambda P_t+V_jQ_{t,j} \nonumber\\
\text{s.t.} \quad &\sum_{k=1}^K\alpha_{i,k}= 1 \nonumber, \quad \forall i,
\end{align}
Now for any given sub-carrier allocation and relay assignment, the optimal power allocation at BS and j-th AR can be obtained from
\begin{align}
\max_{p_i \geq 0 , u_{i,j} \geq0} \ \ &\bigg(\log_2\left(\frac{b_{i,j,k}+a_{i,j}b_{i,j,k}p_i+u_{i,j}b_{i,j,k}c_{i,j}}{c_{i,j}+a_{i,j}c_{i,j}p_i+u_{i,j}b_{i,j,k}c_{i,j}}\right)\label{23}\\&-\lambda p_i-V_ju_{i,j}\bigg)\nonumber.
\end{align}
The problem (\ref{23}) is convex  and closed form solution can be obtained by exploiting the standard techniques similar to section-II-B. Applying KKT conditions we get:
\begin{align}
p_i^*= \left(\frac{-Y_i+\sqrt{Y_i^2-4X_iZ_i}}{2X_i}\right)^+,
\end{align}
where the values of $X_i$, $Y_i$ and $Z_i$ are given in Table. II and 

\begin{table*} [t] 
 \begin{center}
     \caption{}
 \begin{tabular}{|c|c|}
     \hline  $X_i$ & 
    $(a_{i,j}^3b_{i,j,k}^3\lambda V_j-a_{i,j}^3b_{i,j,k}c_{i,j}^2\lambda V_j-a_{i,j}^3b_{i,j,k}^2c_{i,j}\lambda V_j$
   \nonumber\\&\qquad  $+a_{i,j}^3c_{i,j}^3\lambda V_j+a_{i,j}^2b_{i,j,k}^3c_{i,j}\lambda^2 +a_{i,j}^2b_{i,j,k}c_{i,j}^3\lambda^2 -2a_{i,j}^2b_{i,j,k}^2c_{i,j}^2\lambda^2 )$ \\ 
   \hline $Y_i$	& $(2a_{i,j}^3b_{i,j,k}^2+2a_{i,j}^3c_{i,j}^2-4a_{i,j}^3b_{i,j,k}c_{i,j}-2a_{i,j,k}^2b_{i,j,k}^2c_{i,j}\lambda V_j+3a_{i,j}^2b_{i,j,k}^3\lambda V_j-3a_{i,j}^2b_{i,j,k}c_{i,j}^2\lambda V_j+a_{i,j}^2c_{i,j}^3\lambda V_j$
    \nonumber\\&\qquad 
    $+2a_{i,j}b_{i,j,k}^3c_{i,j}\lambda^2+2a_{i,j}b_{i,j,k}c_{i,j}^2\lambda^2-4a_{i,j}b_{i,j,k}^2c_{i,j}^2\lambda^2-a_{i,j}^3b_{i,j,k}^3V_j-3a_{i,j}^3b_{i,j,k}c_{i,j}^2V_j+3a_{i,j,k}^3b_{i,j,k}^2c_{i,j}V_j+a_{i,j,k}c_{i,j}^3V_j)$ \\
     \hline $Z_i$& $(a_{i,j}^2b_{i,j,k}^2+a_{i,j}^2c_{i,j}^2-2a_{i,j}^2b_{i,j,k}c_{i,j}+a_{i,j}b_{i,j,k}^3\lambda  V_j-2a_{i,j}b_{i,j,k}^2c_{i,j}\lambda V_j+a_{i,j}b_{i,j,k}c_{i,j}\lambda V_j-2a_{i,j}b_{i,j,k}c_{i,j}^2\lambda V_j$
     \nonumber\\&\qquad $+b_{i,j,k}^3c_{i,j}\lambda^2+b_{i,j,k}c_{i,j}^3\lambda^2-2b_{i,j,k}^2c_{i,j}^2\lambda^2-a_{i,j}^2b_{i,j,k}^3V_j-3a_{i,j}^2b_{i,j,k}c_{i,j}^2V_j+3a_{i,j}^2b_{i,j,k}^2c_{i,j}V_j+a_{i,j}^2c_{i,j}^3V_j) $ \\ 
   \hline
   \end{tabular} 
   \end{center}
   \end{table*}

\begin{align}
u_{i,j}^* = \left(\frac{\lambda}{V_j}\left(\frac{U_i}{W_i}+\frac{Y_i+\sqrt{Y_i^2-4X_iZ_i}}{2X_i}\right)\right)^+,
\end{align}
with
 $U_i=\left(c_{i,j}-b_{i,j,k}\right)$ and $W_i=\left( b_{i,j,k}a_{i,j,k}-a_{i,j,k}c_{i,j,k}\right)$.
Putting $p_i^*$ and $u_{i,j}^*$ into (\ref{22}) the dual function can be rewritten as
\begin{align}
D(\lambda, V_j)=\max_{(\alpha_{i,k},\beta_{j,k})}  &\sum_{i=1}^N\sum_{j=1}^J\sum_{k=1}^K \alpha_{i,k}\beta_{j,k}(\text{SR}_{i,j,k}^* \nonumber\\&-\lambda p_i^*-V_ju_{i,j}^*) \\ 
\text{s.t.} \quad &\sum_{k=1}^K\alpha_{i,k}= 1 \nonumber, \quad \forall i,
\end{align}
where $\text{SR}_{i,j,k}^*$ is given by:
\begin{align*}
 \text{SR}_{i,j,k}^*=&\log_2\left(\frac{b_{i,j,k}+a_{i,j}b_{i,j,k}p_i^{*}+u_{i,j}^{*}b_{i,j,k}c_{i,j}}{c_{i,j}+a_{i,j}c_{i,j}p_i^{*}+u_{i,j}^{*}b_{i,j,k}c_{i,j}}\right)\\&-\lambda p_i^{*}-V_ju_{i,j}^{*}\nonumber.
\end{align*}
Now, we need to find the optimal sub-carrier allocation and relay assignment. For immediate recovery of the binary variables $\alpha_{i,k}$ and $\beta_{j,k}$, we define a new variable $\pi_{i,j,k} \in \{0,1\}$ such that $\pi_{i,j,k}=1$ if $\alpha_{i,k}\beta_{j,k}=1$ and zero otherwise. The above problem can be rewritten as 
\begin{align}
D(\lambda, V_j)=\max_{\pi_{i,j,k}} &\sum_{i=1}^N\sum_{j=1}^J\sum_{k=1}^K \pi_{i,j,k}(\text{SR}_{i,j,k}^*-\lambda p_i^*-V_ju_{i,j}^*)\\
\text{s.t.}\quad &\sum_{k=1}^K\sum_{j=1}^J\pi_{i,j,k}= 1 \nonumber,\quad \forall i.
\end{align}
The constraint in above optimization ensures that each sub-carrier is assigned to one relay and one user.  
The optimum solution of above problem is to assign a sub-carrier AR pair $(i,j)$ to user $k$ which maximizes the $\text{SR}_{i,j,k}^*$, i.e.,  
\begin{align}
(i^*,j^*,k^*)=\arg \max_{i,j} \ \ \text{SR}_{i,j,k}^*,
\end{align}
Thus, we have
\begin{align}
\pi_{i,j,k}^*&=\begin{cases}
1, &  \text{ for }(i,j,k)=(i^*,j^*,k^*) \\
\\
0, &  \text{otherwise}.
\end{cases}
\end{align}
Now the optimum sub-carrier allocation and relay assignment are obtained. Let $\alpha_{i,k}^*$ and $\beta_{j,k}^*$ denote the optimal assignment variables. Thus, substituting $p_i*$, $u_{i,j}^*$, $\alpha_{i,k}^*$, and $\beta_{j,k}^*$ into (18) we obtain the dual function.  

Next similar to (\ref{10}) and (\ref{11}), we solve the dual problem (\ref{19}) with the sub gradient method \cite{gd}--\cite{9}. The sub gradient updates at $(m+1)$th iteration are:
 
\begin{align*}
\lambda{(m+1)}=\lambda{(m)}+\delta\left(P_t-\sum_{i=1}^N\sum_{j=1}^J\sum_{k=1}^K \alpha_{i,k}\beta_{j,k}p_i \right),
\end{align*}
 \begin{align*}
V_{j}{(m+1)}=V_{j}{(m)}+\delta\left(Q_{t,j}-\sum_{i=1}^N\sum_{k=1}^K \alpha_{i,k}\beta_{j,k}u_{i,j}\right), \forall j.
\end{align*}
In each sub-gradient update, the values of power variables as well as relay selection and sub-carrier assignment are obtained from (20), (21), and (25). The program is terminated at the convergence and the proposed joint optimization algorithm is completed. 


\section{Optimization at the Relay Nodes for Fixed power allocation at BS}\label{s3b}
The previous works in \cite{lat} and \cite{waqas} considered the power optimization at the BS for uniform distribution at the relay. The dynamic relay selection and sub-carrier allocation strategy adopted in this work may assign a relay to multiple users and each relay may have different number of sub-carriers. Thus, the power optimization at each relaying node with independent power constraint becomes more important. In this section, we consider the joint optimization over power allocation at the relaying node, the relay selection and the sub-carrier assignment for the uniform power allocation at the BS i.e., $p_i=PT/N, \forall i$. The corresponding optimization problem can be written as
\begin{align}
\max_{(u_{i,j},\alpha_{i,k},\beta_{j,k})}\quad &\ \sum_{i=1}^N\sum_{j=1}^J\sum_{k=1}^K \alpha_{i,k}\beta_{j,k}\nonumber\\&\log_2\left(\frac{{b_{i,j,k}+a_{i,j}b_{i,j,k}p_i+u_{i,j}b_{i,j,k}c_{i,j}}}{{c_{i,j}+a_{i,j}c_{i,j}p_i+u_{i,j}b_{i,j,k}c_{i,j}}}\right) \label{29}\\
\text{s.t.} \ \ \quad &\sum_{k=1}^K\alpha_{i,k}= 1, \nonumber \quad \forall i,
\\&\sum_{i=1}^N\sum_{k=1}^K \alpha_{i,k}\beta_{j,k}u_{i,j} \leq Q_{t,j} \;\;\forall j. \nonumber
\end{align}
This is a binary integer programming problem. Similar to section-III-B, we adopt dual decomposition approach. For any given relay assignment and sub-carrier allocation, the power optimization at different relay can be obtained by solving following $J$ sub-problems
\begin{align}
\max_{u_{i,j} \geq0}\bigg(\log_2\left(\frac{b_{i,j,k}+a_{i,j}b_{i,j,k}p_i+u_{i,j}b_{i,j,k}c_{i,j}}{c_{i,j}+a_{i,j}c_{i,j}p_i+u_{i,j}b_{i,j,k}c_{i,j}}\right)-\zeta_ju_{i,j}\bigg)\nonumber,
\end{align}
$\forall j\in\{1,2,\ldots,J\}$ and $\zeta_j$ is the Lagrange multiplier corresponding to $j$th relay power constraint.  The resultant value of $u_{i,j}^*$ is given as
\begin{align}
&u_{i,j}^*=\nonumber\\& \!\!\!\left(\!\frac{-2a_{i,j}b_{i,j,k}c_{i,j}^3p_i+\sqrt{(2a_{i,j}b_{i,j,k}c_{i,j}^3p_i)^2-4b_{i,j,k}^2c_{i,j}^2)\Omega_i}}{2b_{i,j,k}^2c_{i,j}^2}\right)^+\!\!,
\end{align}
where $\Omega_i=c_{i,j}^2+(a_{i,j}c_{i,j}p_i)^2+2a_{i,j}c_{i,j}^2p_i-\frac{1}{\zeta_j}\big({b_{i,j,k}c_{i,j}^2}-{a_{i,j}b_{i,j,k}c_{i,j}^2p_i}+{b_{i,j,k}^2c_{i,j}}+{a_{i,j}b_{i,j,k}^2c_{i,j}p_i}\big)$. Now, similar to the previous section, we substitute the value of power variable in the corresponding dual function and the optimal relay assignment and sub-carrier allocation ($\alpha_{i,j}^*, \beta_{j,k}^*$) can be obtained in similar fashion. Finally, the dual problem is solved from the sub-gradient method. The detail steps of the solution are missing for simplicity and are similar to the solution proposed in section III-B.

\section{Power Optimization for Given Subcarrier Allocation and Relay Assignment}\label{s3c}
The solution proposed in Section III and Section IV first find the power allocation for all the possible relay assignment and the sub-carrier allocation and then based on the obtained optimal power optimization, select the best relay selection and sub-carrier assignment. This requires to solve $NJK$ sub-problems in each iteration of the sub-gradient update. In this section, we present a sub-optimal scheme where the joint power allocation at the source and relay is obtained for the predefined $\alpha_{i,j}$ and $\beta_{j,k}$.  The steps involved in the algorithm are listed as follows:
\begin{itemize}
\item[1.] Randomly allocate all the sub-carriers such that the $i$-th sub-carrier is exclusively allocated to a unique user-relay pair $(j,k)$. Thus, both $\alpha_{i,k}$, and $\beta_{j,k}$ are obtained.   
\item[2.]With obtained sub-carrier and AR allocation, the optimization is similar to single user single AR power allocation problem, however with $J+1$ independent power constraints instead of two. i.e.,  
\begin{align}
\max_{p_{i},u_{i,j}}\quad &\ \sum_{i=1}^N\sum_{j=1}^J\sum_{k=1}^K \alpha_{i,k}\beta_{j,k} \text{SR}_{i,j,k} \\
\text{s.t.} \quad  &\sum_{i=1}^N\sum_{j=1}^J\sum_{k=1}^K \alpha_{i,k}\beta_{j,k}p_i \leq P_t, \nonumber
\\ \quad  &\sum_{i=1}^N\sum_{k=1}^K \alpha_{i,k}\beta_{j,k}u_{i,j} \leq Q_{t,j}. \quad \forall j \nonumber
\end{align}
This problem can be solved using similar dual technique in Section III-B. However note that now we need to find only $N$ power variables instead of $NKJ$ variables for power allocation at each step of the dual update.  
\end{itemize}

\section{Simulation Results}\label{sim}
In this section, we present the simulation results to show the performance of the proposed schemes. We choose  6 tap channels taken from i.i.d Guassian random variables for all links and assume same noise variance at all nodes.
\begin{figure}
\includegraphics[width=9.6cm, height=7.2cm]{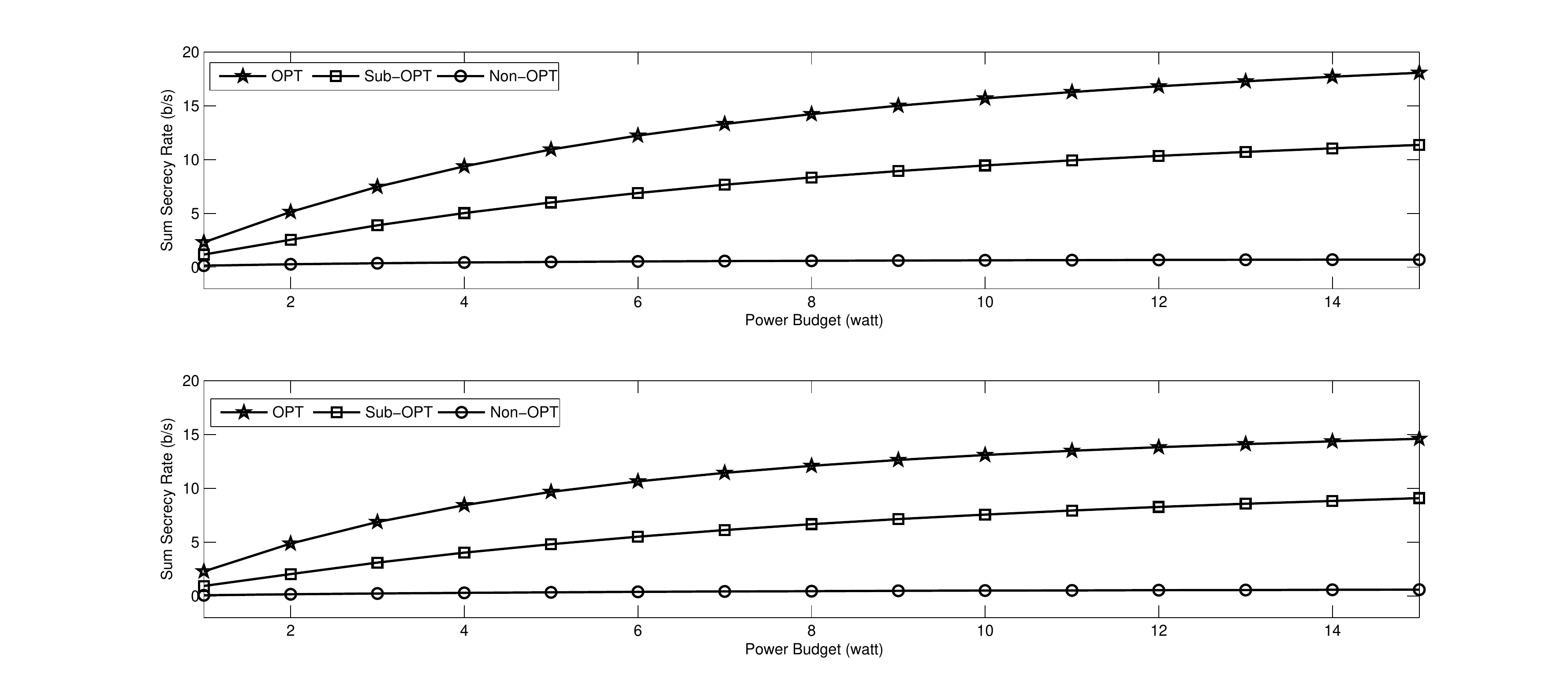}
\caption{Sum Secrecy Rate vs Power Budget}
\label{Fig 3: Sum Secrecy Rate vs Total Transmit Power (PT)}
\end{figure}
For analysis of the results, we compare the performance of following schemes:
\\ \textbf{OPT:} This scheme includes optimization of power loading over all sub-carriers at both BS and AR node for the single relay case, presented in Section-\ref{s2b}.\\ \textbf{Sub-OPT:} In this algorithm, we consider the power optimization at the relay node only, while uniform power distribution is considered at the BS. Thus, it is similar problem as presented in Section IV, however for the single user single relay node case. The step wise detail of the scheme is missing due to simplicity.\\
\textbf{J-OPT:} This refers to the joint optimization of  relay selection, sub-carrier allocation, and power allocation at all transmitting nodes, presented in Section-\ref{jnt}.\\
\textbf{Sub-OPT-I:} It represents the solution presented in Section-\ref{s3b}.\\
\textbf{Sub-OPT-II:} The solution with power optimization at the BS and all the relay nodes under fixed relay assignment and sub-carrier allocation, as given in Section-\ref{s3c}.\\
\textbf{Non-OPT:} This refers to the case with fixed sub-carrier allocation, predefined relay selection, and equal power distribution among sub-carriers at each transmitting nodes. Hence, for single user single relay case, this corresponds to uniform  power allocation among all sub-carriers at the two nodes.

Figure 3 presents the results for single relay case where y-axis represents sum secrecy rate and x-axis represents total power budget. Same power budget is considered at BS and AR, while we have  set N=64 and N=32 for the upper and  the lower subplots, respectively. It can be clearly noted that OPT scheme  outperforms the remaining two schemes and the Sub-OPT performs better than the Non-OPT as presented in Fig. 3. The performance gap between OPT and other candidates increases with the increase in number of sub-carriers and power budget. The better performance with increasing the number of sub-carriers is due to the higher degree of freedom in power allocation. The increase in the power budget not only increase the sum secrecy rate for both OPT and Sub-OPT but also increases the gap. This is because of the fact that the OPT scheme efficiently distributes the available power budget among different sub-carriers at the two transmitting nodes while the Sub-OPT allocates power uniformly among sub-carriers at the BS. Non-OPT does not
provide secure communication as this scheme has zero sum secrecy rate i.e., the feasible solution does not exist with uniform power allocation. Hence, the resource optimization is mandatory for providing secure communication at the physical layer. \\
\begin{figure}
\includegraphics[width=9.6cm, height=7.2cm]{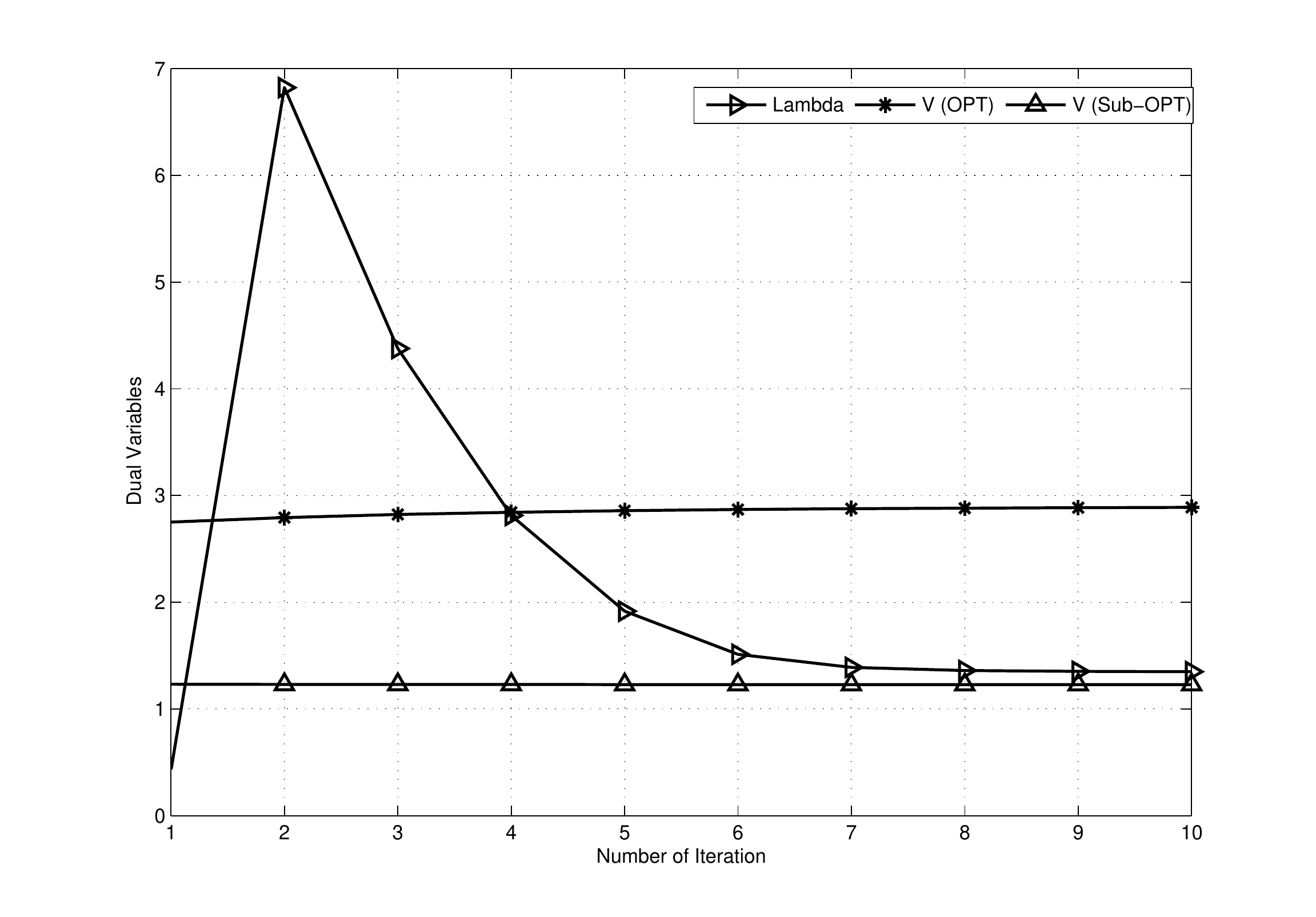}
\caption{ Convergence Rate}
\label{Fig 3: Sum Secrecy Rate vs Total Transmit Power (PT)}
\end{figure}
Next, in Fig. 4 we show the convergence behavior of the dual variables for the two optimization schemes. Please note that, the OPT involve two dual variables while the Sub-OPT has a single dual variable. It can be observed that both the schemes converge within acceptable number of iterations. Further, it is noted that the OPT provides higher performance at the cost of few more number of iterations for convergence. On the other hand, the Sub-OPT provides much better performance over the Non-OPT without requiring high burden of time consumption in terms of number of iterations.  

In Fig. 5, we consider multiuser multi-relay scenario with J=4 and K=12. Similar to Fig. 3, same available power budget is assumed at all nodes and results are obtained with N=32 as well as for N=64. It can be seen from Fig. 5 that there is a clear gap between the J-OPT and the other schemes while Sub-OPT-I outperforms Sub-OPT-II. Increasing the number of sub-carriers increases the degree of flexibility of power allocation which results in increasing secrecy rate of all schemes. It is also interesting to note that the enhancement in performance of J-OPT scheme with increasing N is higher than the other schemes.
 
For Fig. 6 and Fig. 7 we have taken a single realization of Gaussian random channels to show the possible effects of adding a new relay or user in the system. Figure 6 shows the impact of varying number of relays from 1 to 4, with $P_t=Q_t=7$ , K=12 and N=64. We can see similar trends as in Fig, 5. Increasing the number of relays provides enhanced performance for all schemes. However, the percentage increase of the sum secrecy rate of J-OPT and Sub-OPT-1 is much more than the other two players. This is because, both of these schemes involve optimizing relay selection and sub-carrier assignment while the other two use fixed. Last but not least, the rate of increase in secrecy rate is more from J=1 to J=3 and becomes a bit low from J=3 to J=4. This is because, initial addition of relays provide a higher freedom in resource allocation. For a more closer look into exact values, the results are also depicted in Table III. 
\begin{table}
     \caption{Sum secrecy rate for different number of relays}
 \begin{tabular}{|c|c|c|c|c|}
   \hline
    Number of Relays & J-OPT & Sub-OPT I & Sub-OPT II & Non-OPT \\
   \hline
    $1$	& $18.10$	& $7.01$	& $2.94$	& $0$	\\
    \hline
    $2$	& $28.17$	& $13.88$	& $3.97$	& $0$	\\
   \hline
   $3$	& $39.47$	& $19.01$	& $3.99$	& $0$	\\
    \hline
     $4$	& $40.97$	& $19.10$	& $4.73$	& $0$	\\
    \hline
   \end{tabular} 
   \end{table}

Finally, to complete the analysis, we check the sum secrecy rate for different number of users with fixed number of relays. The results are plotted in Fig. 7 with N=64, PT=QT=7, and J=4. Again, superiority of J-OPT over all competing candidates is clear. In J-OPT increasing the number of users enhances the performance if the channel gains of new user are better compared to old users.  This is due to the fact that higher number of users provide better channel conditions and higher flexibility in power allocation. The J-OPT considers all parameters jointly, hence, we observe a significant performance gap increase with increasing the number of users. 

\begin{figure}
\includegraphics[width=9.6cm, height=7.2cm]{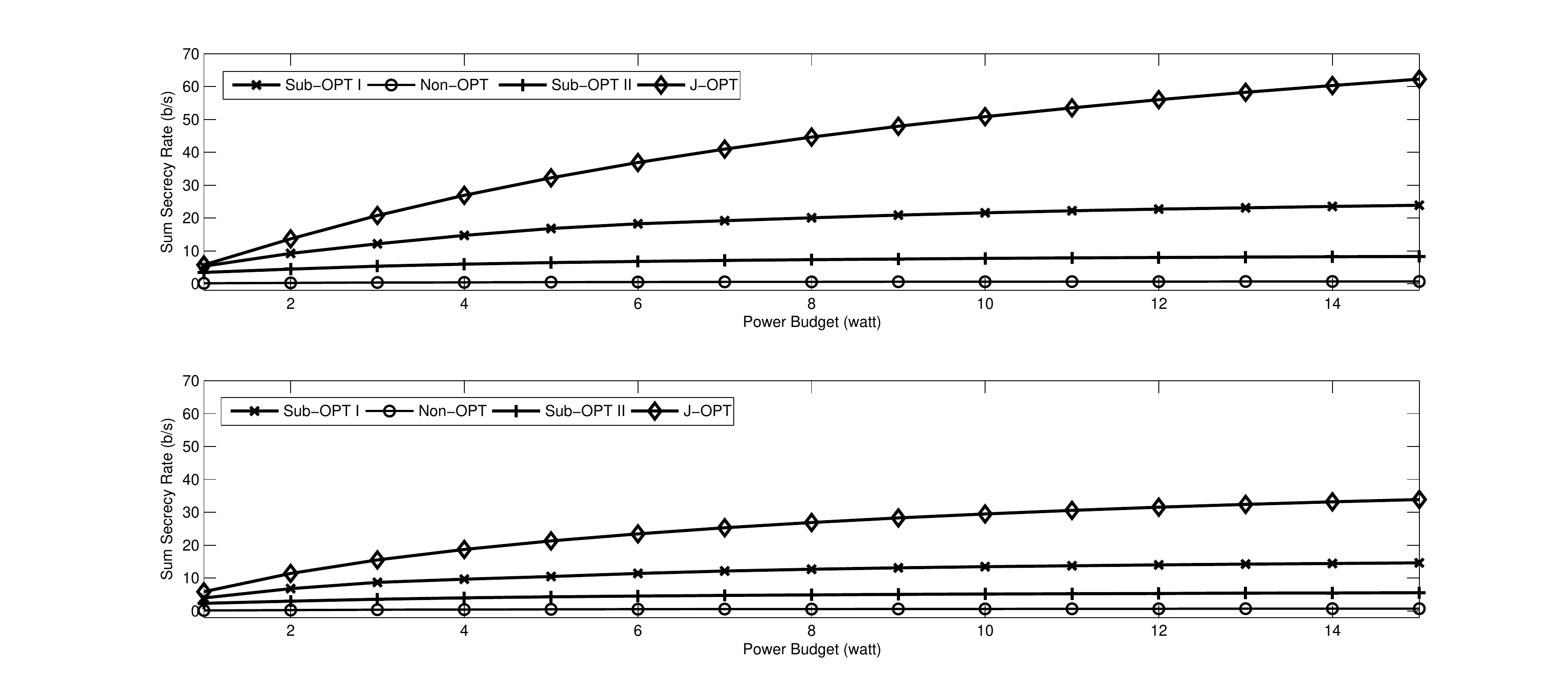}
\caption{Sum Secrecy Rate vs Power Budget}
\label{Fig 3: Sum Secrecy Rate vs Total Transmit Power (PT)}
\end{figure}
\begin{figure}
\includegraphics[width=9.6cm, height=7.2cm]{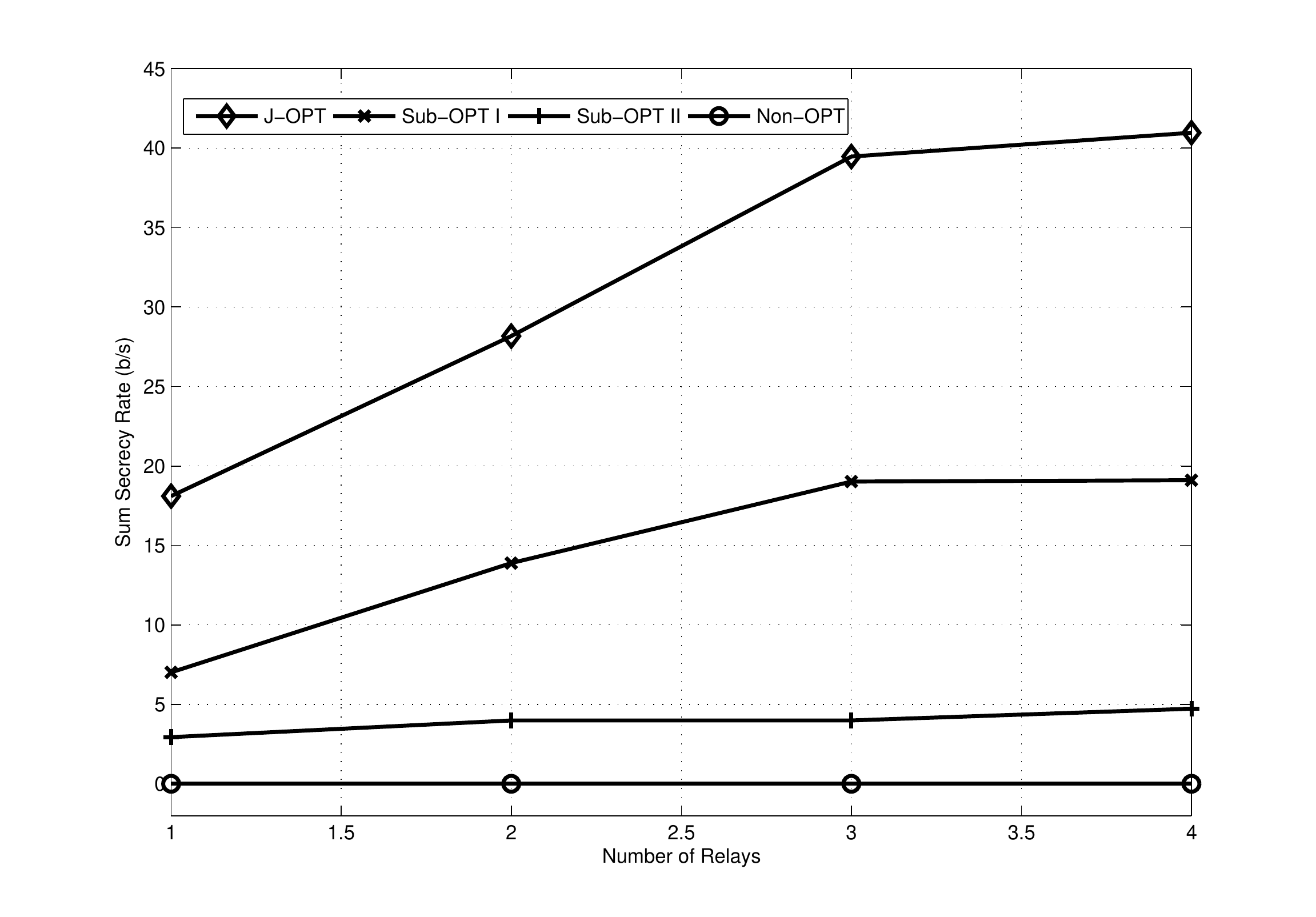}
\caption{Sum Secrecy Rate vs Number of Relays}
\label{Fig 3: Sum Secrecy Rate vs Total Transmit Power (PT)}
\end{figure}

\begin{figure}
\includegraphics[width=9.6cm, height=7.2cm]{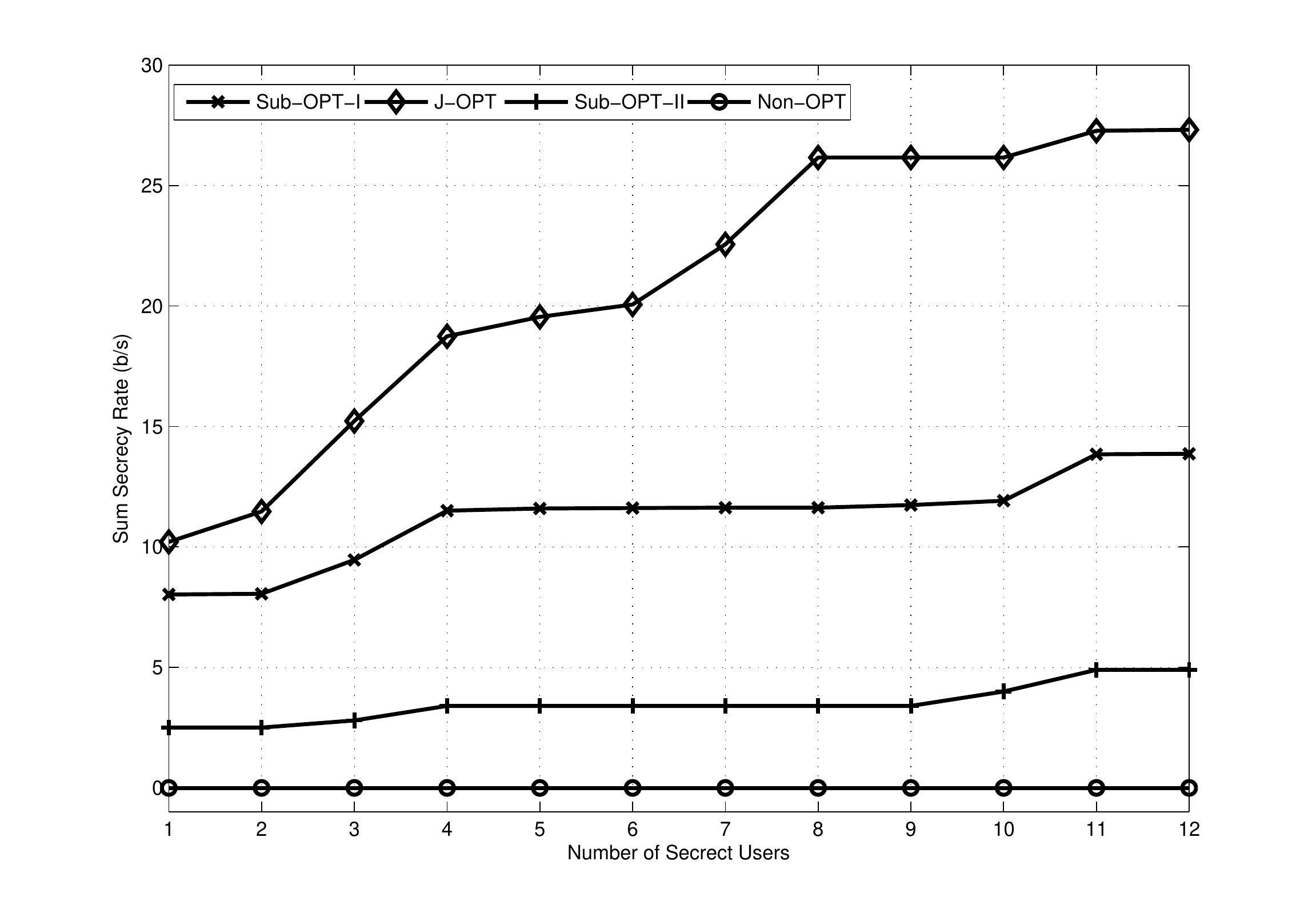}
\caption{Sum Secrecy Rate vs Number of Users}
\label{Fig 3: Sum Secrecy Rate vs Total Transmit Power (PT)}
\end{figure}
\section{Conclusion}\label{con}
This paper considered resources allocation problem to enhance PLS in AF relay assisted wireless networks. Joint optimization problem of  power allocation at different transmitting nodes, relay assignment and the sub-carrier allocation was studied. For practical reasons, separate power constraint was considered at  the BS and each relaying node. A dual decomposition framework was  adopted to find an efficient solution for sub-carriers allocation, relays assignment and power loading over all sub-carriers. The target was to maximize the sum secrecy rate of the system. Further, sub-optimal schemes were also presented. Simulation results validated the performance of all proposed schemes. Joint optimization and sub-optimal schemes outperformed the trivial solutions. It was observed that the gain of the proposed joint optimization solution increases with the increases in the number of sub-carriers, number of relays and total power budget.

\end{document}